# On the optimal allocation of assets in investment portfolio with application of modern portfolio management and nonlinear dynamic chaos theories in investment, commercial and central banks

Dimitri O. Ledenyov and Viktor O. Ledenyov

*Abstract* – The investment economy is a main characteristic of prosperous society. The investment portfolio management is a main financial problem, which has to be solved by the investment, commercial and central banks with the application of modern portfolio theory in the investment economy. We use the learning analytics together with the integrative creative imperative intelligent conceptual co-lateral adaptive thinking with the purpose to advance our scientific knowledge on the diversified investment portfolio management in the nonlinear dynamic financial system. We apply the econophysics principles and the econometrics methods with the aim to find the solution to the problem of the optimal allocation of assets in the investment portfolio, using the advanced risk management techniques with the efficient frontier modeling in agreement with the modern portfolio theory and using the stability management techniques with the dynamic regimes modeling on the bifurcation diagram in agreement with the dynamic chaos theory. We show that the bifurcation diagram, created with the use of the logistic function in Matlab, can provide some valuable information on the stability of combining risky investments in the investment portfolio, solving the problem of optimization of assets allocation in the investment portfolio. We propose the Ledenyov investment portfolio theorem, based on the Lyapunov stability criteria, with the aim to create the optimized investment portfolio with the uncorrelated diversified assets, which can deliver the increased expected returns to the institutional and private investors in the nonlinear dynamic financial system in the frames of investment economy.

The *investment economy* is a main characteristic of prosperous society. The *investment economy* is researched in the *science of economics*, which considers the research problems of wealth creation and distribution in the social systems. The accurate definition of the science of economics is made in *Ueda (1904), Mano (1968-1969)*:

"*Economics* in wide sense:

*A*. *Political Economy* (a study of economics from the standpoint of the social economic system).

*B*. *Science of Business Administration* (a study of economics from the standpoint of the economic unit):

*a*. *Science of Finance* (a study of the management of national economy);

*b*. *Housekeeping* (a study of home economics);

*c*. *Science of Commerce* (a study of economics of a business enterprise)."

At later date, *Ueda (1937)* more accurately described the relations between the political economy and the science of business administration in *Mano (1968-1969)*:

"*Economics* in a wide sense:

*A*. *Political Economy* or Economics (The main object of this science is the price phenomenon).

*B*. *Science of Business Administration* in a wide sense (The science of economic organization, which has an object):

*a*. *Science of Finance* (The science of state economy);

*b*. *Housekeeping* (The science of home economy);

*c*. *Science of Business Administration* (The science of business economy)."

In this research paper, we make the innovative research in the *science of finance*, which is a part of the *science of business administration* under the general framework of the *science of economics*. In the science of finance, we will focus on the *investment portfolio* problem, considering the various topics on the investment portfolio optimization. In our time, the research in the modern science of finance is done with the use of the financial mathematics (the *econometrics*), which began to develop in the beginning of last century in *Bachelier (1900)*. *Merton (2001)* writes: "The origins of much of the mathematics in modern finance can be traced to *Louis Bachelier's 1900 dissertation* on the theory of speculation, framed as an option-pricing problem. This work marks the twin births of both the continuous-time mathematics of stochastic processes and the continuous-time economics of derivative-security pricing. In solving his option-pricing problem, *Bachelier* provides two different derivations of the classic partial differential equation for the probability density of what later was called a *Wiener* process or *Brownian* motion process. In one derivation, he writes down a version of what is now commonly called the *Chapman-Kolmogorov* convolution probability integral in one of the earliest examples of that integral in print. In the other, he uses a limit argument applied to a discrete-time binomial process to derive the continuous-time transition probabilities. *Bachelier* also develops the method of images (or reflection) to solve for a probability function of a diffusion process with an absorbing barrier. All this in his thesis five years before the publication of *Einstein's* mathematical theory of *Brownian motion*." *Shiryaev (1998)* explains: "*Bachelier (1900)* certainly was the first researcher, who, with the aim to describe the dynamics of shares prices, used the model of "random walks and their limiting formations," which, speaking by the modern language, represent nothing other than the *Brownian motion*." The research in the modern science of finance is also frequently done with the application of financial physics (the *econophysics*), which has been developed over a long time period. It makes sense to explain that the *econophysics*, which aims to understand the nature of complex processes in the economics and finances, using the knowledge base with the theories, experiments and computer modeling in the field of physics, has a long history in the sciences of fundamental and applied economics and finances. Over the recent decades, many scientists made their significant research contributions to the **science of econophysics** as described in *Mantegna, Stanley (1999)*, *Ilinski (2001)*, *Bouchaud, Potters (2003)*, *Sornette (2003)*, *Sinha, Chatterje, Chakraborti, Chakrabarti (2010)*. Presently, there is a strong research interest in the *econophysics* from the side of various research groups at the leading universities around the World, which make numerous attempts to re-define the research boundaries of the *science of econophysics*. The most recent definitions of scientific meaning of *econophysics* were proposed in *Yakovenko, Rosser (2009)*, *Chakrabarti B K, Chakrabarti A (2010)*, *Aoyama, Fujiwara, Iyetomi, Sato (2012)*, aiming to explain the essence of research approaches, used in the *econophysics*. *Yakovenko, Rosser (2009)* write: "**Econophysics** is a new interdisciplinary research field, applying methods of statistical physics to problems in economics and finance." *Chakrabarti B K, Chakrabarti A (2010)* explain: "**Econophysics** is a new research field, which makes an attempt to bring economics in the fold of natural sciences or specifically attempts for a "physics of economics". *Aoyama, Fujiwara, Iyetomi, Sato (2012)* state: "**Econophysics** is a relatively new discipline that applies concepts and methodologies that originated in physics to the problem of obtaining a better understanding of a wide variety of complex phenomena of a socio-econo-techno nature, including economic, sociological and financial behavior."

*Chakrabarti B K, Chakrabarti A (2010)* compiled a list of most important research results and achievements in the field of econophysics:

"**(i)** ***Empirical characterization, analyses and modeling of financial markets*** – in particular, the deviation from Gaussian statistics has been established, following the early observations of Mandelbrot and Fama (1960s) – beginning with the studies in 1990s by the groups of *Stanley, Mantegna, Bouchaud, Farmer and others*.

**(ii)** ***Network models and characterization of market correlations among different stocks/sectors*** by the groups of *Mantegna, Marsili, Kertesz, Kaski, Iori, Sinha and others*.

**(iii)** ***Determination of the income or wealth distributions in societies, and the development of statistical physics models*** by the groups of Redner, Souma, Yakovenko, Chakrabarti, Chakraborti, Richmond, Patriarca, Toscani and others. The kinetic exchange models of markets have now been firmly established; this gained a stronger footing with the equivalence of the maximization principles of entropy (physics) and utility (economics) shown by the group of Chakrabarti.

**(iv)** ***Development of behavioral models, and analyses of market bubbles and crashes*** by the groups of *Bouchaud, Lux, Stauffer, Gallegati, Sornette, Kaizoji and others*.

**(v)** ***Learning in multi-agent game models and the development of Minority Game models*** by the groups of *Zhang, Marsili, Savit, Kaski and others*, and the ***optimal resource utilization "Kolkata Paise Restaurant" (KPR) model*** by the group of *Chakrabarti*."

In this research article, we aim to make our new innovative research in the finances, applying the ***econophysics principles*** to complement the existing ***econometrics models*** with the aim to find a solution to the problem of the optimal allocation of assets in the *investment portfolio*, using the advanced risk management techniques with the efficient frontier modeling in agreement with the modern portfolio theory and using the stability management techniques with the dynamic regimes modeling on the bifurcation diagram in agreement with the dynamic chaos theory, in the conditions of ***nonlinear dynamical financial system*** in the ***free market economy***, governed by the ***modern capitalism***. Let us clarify that we will assume that the ***capitalism*** is defined as in *Macmillan Dictionary of Modern Economics (1986), Scott (2007)*: "Political, social, and economic system in which property, including capital assets, is owned and controlled for the most part by private persons. Under capitalism, the price mechanism is used as a signaling system which allocates resources between uses. The extent to which the price mechanism is used, the degree of competitiveness in markets, and the level of government intervention distinguish exact forms of capitalism." It makes sense to add that the detailed characteristics of the *free market economy*, including the valuation and taxation problems in the states with the free market economy under the capitalism, with the particular focus on the transition to and formulation of the ideal free market economy principles were researched in *Ledenyov V O, Clancy R (1990-1993), Ledenyov V O, Foldvary F (1993-1994)*. The authors came to the understanding that all the free market economies consist of a ***market*** (voluntary economic acts) and ***interventions*** (coercive acts imposed on a market by the governments). Going from the *econophysical* analysis, the following practical steps to reduce the intervention and establish the optimal ideal free market economy were proposed in *Ledenyov V O, Foldvary F (1993-1994), Ledenyov V O (1995)*:

**1)** Legalize, preferably through constitutional provisions, voluntary production, exchange (trade), and consumption.

**2)** Set up a registry of all real estate and titleholders.

**3)** Assess the ground rent for all land, including the economic rent of minerals, fishing areas, etc.

**4)** Initiate the collection of the ground rent, gradually increasing it to the value of the economic rent. Preferably, there should be a constitutional provision stating that ground rent is community property.

**5)** At the same time as #4, reduce taxes on wages, income, profits, turnover, sales, exports, and imports (the order depending on political feasibility).

**6)** Initiate effluent (pollution) fees proportional to the damage caused by pollution (Germany provides a model).

**7)** Raise the pollution fee while reducing other taxes (as with #5) over some set time interval until all taxes on production are eliminated, with a constitutional provision banning them.

**8)** Eliminate all remaining economic restrictions, prohibiting only force, fraud, and other acts of coercive and invasive harm. This includes gradually reducing and eliminating all tariffs and trade quotas, as well as restrictions on money and banking, and a constitutional provision protecting freedom of commerce as well as complete personal liberty.

In addition, it makes sense to note that we conduct our research, assuming that the *nonlinear dynamic financial system* is greatly affected by the process of ***globalization***. The *globalization* means the integration of economic activities, across borders, through markets; resulting in the free movement of goods, services, labour and capital in the global single free market in *Wolf (2004, 2006)*. The *globalization* represents the closer integration of the countries and people of the world, which has been brought about by the enormous reduction of costs of transportation and communication, and the breaking down of artificial barriers to the flows of goods, services, capital, knowledge, and (to a lesser extent) people across borders in *Stiglitz (2002)*. The *globalization* has many dimensions such as the globalization of ideas, of knowledge, of civil society in *Stiglitz (2002)*. In the conditions of *globalization*, the well designed, perfectly optimized and efficiently operated financial system represents a main foundation of prosperous developed society with the free market economy in *Ledenyov D O, Ledenyov V O (2012)*. At the same time, we must understand that the presently established global financial and economic systems have a number of functional challenges and proved to be ineffective, resulting in a conclusion that the states with the *free market economies* may also be frequently exposed to the economic stagnation and recession processes as discussed in *Ledenyov D O, Ledenyov V O (2012)*.

Let us continue our investigation with the focus on the main subject of our research, which is the problem of the optimal allocation of assets in the *investment portfolio* in the *nonlinear dynamic financial system* in the frames of *investment economy* in the conditions of globalization. As it was already mentioned, the research on the investment portfolio can only be done with the use of the financial mathematics. Initially, the dynamics of continuously traded investment portfolios was developed with the use of the calculus as proposed by *Kiyoshi Itô*. *Merton (2001)* writes on the theoretical proposals by *Kiyoshi Itô*: "On the mathematical side, *Kiyoshi Itô* was greatly influenced by *Bachelier's* work in his development in the *1940s* and early *1950s* of his stochastic calculus, later to become an essential tool in finance. ... Initially, *Ito's* mathematics found its way into finance with the development of **the continuous-time theory of optimal lifetime consumption and portfolio selection**. This theory used **diffusion processes** to model asset price movements and applied the *Ito* calculus to analyze the dynamics of continuously traded portfolios. The connection between *Itô's* work and *option pricing* was made, when that same continuous trading portfolio modeling tool was used to derive dynamic portfolio strategies that replicate the payoffs to an option, from which the famous *Black-Scholes option pricing theory* was born." Let us explain that, presently, the investment, commercial, central banks usually make their investments by creating the **investment portfolios** with the different **assets**. In the general case, it is possible to characterize the investments by their **expected return** and **standard deviation of return**. The *investment portfolio* building in the process of wealth management is a challenging task, because the capital markets represent the open, non-linear and complex financial systems with the positive or negative feedback loop mechanisms, resulting in the assets prices fluctuations in *Mosekilde (1996, 1996-1997)*, *Beinhocker (2006), Mandelbrot (2004)*. The **Modern Portfolio Theory (MPT)** was proposed in *Markowitz (1952, 1956, 1959, 1987)*. The fundamental concept of the *MPT* is that the price changes by the different interrelated assets must be taken to the account during the investment portfolio building. The ***Efficient Frontier (EF)*** concept in the *MPT* was introduced in *Markowitz (1952)*.

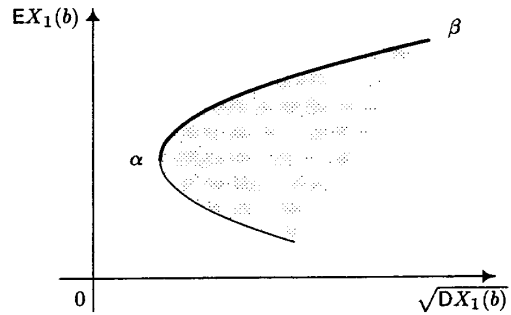

***Fig. 1.*** *Illustration of mean variance analysis by Markowitz (after Shiryaev (1998)).*

*Markowitz (1952)* considers the two characteristics of capital $X_1(b)$:

1) $EX_1(b)$ is the mathematical expectation;

2) $DX_1(b)$ is the dispersion.

In agreement with the *mean variance analysis* by *Markowitz (1952)*, the investment portfolios with a set of points $(EX_1(b), \sqrt{DX_1(b)})$ between the point $\alpha$ and the point $\beta$ on the *efficient frontier* curve have the maximum mean value of capital at the minimum value of dispersion in Fig. 1 in *Shiryaev (1998)*.

In Fig. 2, the illustration of the efficient frontier of all the investments is shown by *Hull (2005-2006, 2010, 2012)*.

$$\mu_P = w_1\mu_1 + w_2\mu_2 \qquad \sigma_P = \sqrt{w_1^2\sigma_1^2 + w_2^2\sigma_2^2 + 2\rho w_1 w_2 \sigma_1 \sigma_2}$$

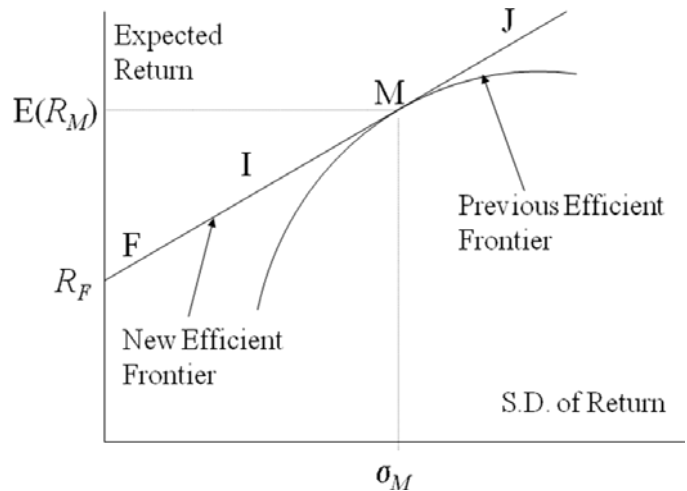

***Fig. 2.*** *Efficient frontier of all investments (after Hull (2005-2006, 2010, 2012)).*

As explained in *Wikipedia (2012)*, the combination of risky assets (without the risk-free assets) can be plotted in the *risk-expected return space*, and the collection of all such possible portfolios defines a region in this space. The upward-sloped (positively-sloped) part of the left boundary of this region, a hyperbola, is then called the "***efficient frontier***". The efficient frontier is then the portion of the opportunity set that offers the highest expected return for a given level of risk, and lies at the top of the opportunity set or the feasible set. A combination of risky assets, i.e. a portfolio, is referred to as "efficient," if it has the best possible expected level of return for its level of risk (usually proxied by the standard deviation of the portfolio's return) in Fig. 3.

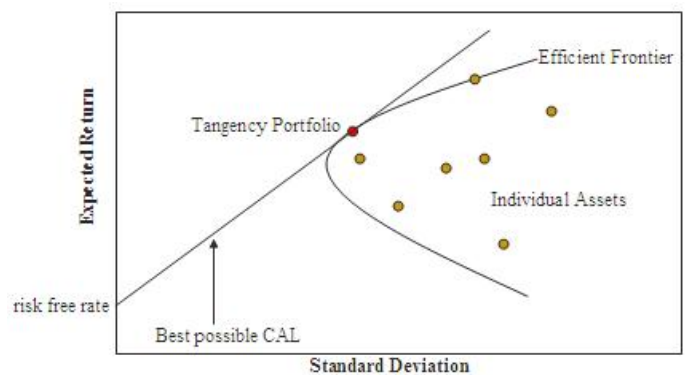

*Fig. 3. The efficient frontier (after Wikipedia (2012)).*

The research contributions to the *MPT* were comprehensively analyzed by *Kaplan (1998)*, who explains: "*Markowitz* extends the techniques of *linear programming* to develop the *critical line algorithm*. The **critical line algorithm** identifies all the feasible portfolios that minimize risk (as measured by variance or standard deviation) for a given level of expected return and maximize expected return for a given level of risk. When graphed in **standard deviation** versus **expected return** space, these portfolios form the **efficient frontier**. The efficient frontier represents the trade-off between risk and expected return faced by an investor, when forming his portfolio." *Kaplan (1998)* highlights: "*Markowitz* showed that while there are infinitely many efficient portfolios, you only need a limited number of *corner portfolios* to identify all *efficient portfolios*. Not all *efficient portfolios* contain all assets. Moving along the *efficient frontier*, a *corner portfolio* is located, where an asset weight or slack variable is either added or dropped. Every *efficient portfolio* is a linear combination of the two *corner portfolios* immediately adjacent to it. Thus, by locating all *corner portfolios*, the *critical line algorithm* generates the entire *efficient frontier*." In addition, *Kaplan (1998)* writes: "In general, the maximizing expected utility of ending period wealth by choosing portfolio weights is a complicated stochastic nonlinear programming problem. Markowitz asserted that, if the utility function can be approximated closely enough by a second-order *Taylor expansion* over a wide range of returns, then expected utility will be approximately equal to a function of expected value (mean) and variance of returns. This allows the investor's problem to be restated as a **mean-variance optimization problem** so that the objective function is a quadratic function of portfolio weights." *Engle (2003, 2006)* states: "*Markowitz (1952)* and *Tobin (1958)* associated risk with the variance in the value of a portfolio." The **Tobin's mutual fund theorem** says that the investment portfolio's assets allocation problem can be viewed as a decision to allocate between a *riskless asset* and a *risky portfolio* in *Tobin (1958)*. In the mean-variance approach framework, the cash can serve as a proxy for a *riskless asset;* and an *efficient portfolio* on the *efficient frontier* serves as the risky portfolio, such, that any allocation between the cash and this portfolio dominates all other portfolios on the *efficient frontier*. This portfolio is called a **tangency portfolio**, because it is located at the point on the *efficient frontier*, where a tangent line that originates at the riskless asset touches the efficient frontier. *Engle (2003, 2006) explains:* "*Sharpe (1964)* developed the implications, when all investors follow the same objectives with the same information. This theory is called the **Capital Asset Pricing Model** or **CAPM**, and shows that there is a natural relation between expected returns and variance". It is necessary to state that the **Sharpe ratio** is a measure of *return-to-risk* that plays an important role in the *investment portfolio* analysis and it was introduced in *Sharpe (1966)*. Specifically, the *investment portfolio* that maximizes the *Sharpe ratio* is also the **tangency portfolio** on the *efficient frontier* from the *mutual fund theorem* in *Sharpe, Alexander, Bailey (1999)*. The maximum *Sharpe ratio investment portfolio* is situated on the *efficient frontier*. Summarizing all the important research results on the *MPT*, it is necessary to say that the most significant research achievements in the *MPT* were presented in *Markowitz (1952, 1956, 1959, 1987), Tobin (1958), Sharpe (1964, 1966),* and *Merton (1969, 1970, 1971, 1972, 1973a, 1973b, 1977a, 1977b, 1982, 1983a, 1983b, 1990, 1992, 1993a, 1993b, 1994, 1995a, 1995b, 1997, 1998, 1999, 2001)*.

Now, let us shortly consider the various *assets*, which can be used in the *investment portfolio*:

*1) Credit Derivatives;*

*2) Equity Derivatives;*

*3) Other Derivatives*.

The definition of derivatives is presented in *Hull (2005-2006, 2010, 2012)*: "A **Derivative** is an instrument whose value depends on the values of other more basic underlying variables." In the *investment portfolio*, the *derivatives* can be used in a number of ways in *Hull (2005-2006, 2010, 2012)*:

**1.** To *hedge risks*;

**2.** To *speculate* (take a view on the future direction of the market);

**3.** To *lock in an arbitrage profit*;

**4.** To *change the nature of a liability*;

**5.** To *change the nature of an investment* without incurring the costs of selling one portfolio and buying another.

*Ledenyov V O, Ledenyov D O (2012)* explained that: "The **Structured Investment Vehicles** (*SIV*), which include the structured credit products, are extremely complex. The purpose of *structured credit products* is to give fixed income investors fully rated and leveraged exposure to the main **Credit Derivatives Indices**. The **investment portfolio** building with the positions hedged by the different means depends on the investor's expertise in the following investment vehicles in *Ledenyov V O, Ledenyov D O (2012)*:

*1. <u>Credit Derivatives</u>* (*CD* are the financial contracts whose payoffs explicitly depend on the behaviour of one or more indices): Collateralized Debt Obligations (*CDO*), Constant Proportion Debt Obligations (*CPDO*), and investment protection mechanisms such as the Synthetic Collateralized Debt Obligations (*SCDO*), Credit Default Swaps (*CDS*), Credit Default Swap Index (*CDSI*), Loan only Credit Default Swaps (*LCDS*), Credit Default Swaps of *ABS* (*ABCDS*), Variance Swaps (*VS*), Constant Proportion Portfolio Insurance (*CPPI*), Contracts for Difference (*CFD*);

*2. <u>Equity Derivatives</u>*: Futures, Asian Options, Barrier Options, Compound Options, Look-back Options, Vanilla Stock Options (put and call options), Vix Options;

*3. <u>Other Derivatives:</u>* Interest Rate Swaps."

Let us more comprehensively consider the **Credit Derivatives (CD)** and **Equity Derivatives**, using the recent research findings in the *Financial Times (2005-2012)*:

*1. <u>Credit Derivatives (CD)</u>* are defined as the investment vehicles that are based on the **Corporate Bonds** and give their owners protection against a default. It is necessary to comment that the **Corporate Bonds** are secured against the issuers balance sheet or the **Covered Bonds** are secured against the pools of **Mortgages** or public-sector **Loans**, making the corporate and covered bonds one of the safest, highest rated issues in the fixed-income markets. The investment banks pool corporate bonds for sale, for example, in the form of the **Collateralized Debt Obligations (CDO)**, and sell off pieces of the pool. Structure of pool consists of both the riskiest slice and the conservative slice. The high default correlation and law default correlation slices are distinguished. The asset backed or **Structured Finance (SF)** backed paper makes up the vast majority of *CDO* issuance. The *CDO* concept actually contributed a lot to the **Capital Market** efficiency, however it is necessary to note that the *CDO* based business model needs greater levels of protection and more pragmatic pricing. The concepts of diversification and collateralization are sound - provided they are combined with some judgment, however the substituting collectivization and collateralization for credit analysis really does not work and merely exacerbates the market's innate tendency to misallocate capital during the more enthusiastic parts of the cycle. The *Synthetic Collateralized Debt Obligations (Synthetic CDO)* replaces the pools of bonds – specifically, with a type called *Credit Default Swaps (CDS)*. The swaps are like insurance policies. They insure against a bond default. Owners of bonds can buy *CDS* on their bonds to protect themselves. *CDS* has improved liquidity in credit generally, much as swaps did in the pure rate markets. The ***derivative products*** allow investors to tranche risk very finely, to bundle risk, but they also allow lots of new entrants to financial markets that otherwise wouldn't be there, because the barriers to entry have been lowered so much. The transparent ***Price Discovery*** process and ability to access ***Secondary Market Pricing*** quickly and easily contribute to the transparency in the *CDO* and *CDS* markets, and is considered as a cornerstone of confidence in financial markets in general.

*2. Equity Derivatives*, let share a few thoughts about the following equity derivatives:

*1. Futures Contracts*;

*2. Forward Contracts*;

*3. Options.*

The clear definition of ***Futures Contract*** is made in *Hull (2005-2006, 2010, 2012)*: "A ***Futures Contract*** is an agreement to buy or sell an asset at a certain time in the future for a certain price. By contrast in a spot contract there is an agreement to buy or sell the asset immediately (or within a very short period of time). The ***Futures Prices*** for a particular contract is the price at which you agree to buy or sell. It is determined by supply and demand in the same way as a spot price. Traditionally, the futures contracts have been traded using the open outcry system, where traders physically meet on the floor of the exchange. Increasingly, this is being replaced by electronic trading, where a computer matches the buyers and sellers."

Speaking about the ***Options***, let us make the following definitions:

*1.* A ***Call Option*** is an option to buy a certain asset by a certain date for a certain price (the **Strike Price**);

*2.* A ***Put Option*** is an option to sell a certain asset by a certain date for a certain price (the ***Strike Price***);

*3.* A ***Stellate Option*** is a double option to buy or sell a certain asset by a certain date for a certain price (the ***Strike Price***).

It is necessary to distinguish the ***American Options*** vs. ***European Options***:

*1.* An ***American option*** can be exercised at any time during its life;

*2.* A ***European option*** can be exercised only at maturity.

Also, it is necessary to remember the distinction between the ***Options*** vs. ***Futures/Forwards Contracts***:

***1. Futures/Forward Contract*** gives the holder the obligation to buy/sell at a certain price;

***2. Option*** gives the holder the right to buy or sell at a certain price.

Let us also explain the meaning of the *volatility*: "**Volatility** is the standard deviation of the continuously compounded rate of return in certain period of time," as it is explained in *Hull (2005-2006, 2010, 2012)*. It is possible to form a portfolio, consisting of the stock and the option, which eliminates this source of ***uncertainty***. The ***implied volatility of an option*** is the volatility for which the ***Black-Scholes price*** equals to the *market price*. Traders and brokers often quote ***implied volatilities*** rather than ***dollar prices***. After the options have been issued, it is not necessary to take account of dilution, when they are valued. Before they are issued, we can calculate the cost of each option as $N/(N+M)$ times the price of a regular option with the same terms, where $N$ is the number of existing shares and $M$ is the number of new shares that will be created, if the exercise takes place. The ***high volatility*** in capital markets may lead to the certain financial problems, for example the *liquidity crisis*. Let us mention that the *Black and Scholes (1972, 1973), Black (1989)* and *Merton (1973)* developed a model to evaluate the *pricing of options* and derived the necessary formulas. The examples of the ***risk transfer***, applying the ***diversification***, ***hedging*** and ***insurance***, are described in *Gray, Merton, Bodie (2007)*, and also studied in other research papers in *Riedl, Serafeim (2001)*, *Campbell, Chan, Viceira (2002)*, *Campbell, Serfaty-de Medeiros, Viceira (2007)*:

- *Asset Diversification in Banking Sector*.
- *Equity Swaps as a Method of Diversifying Internationally*.
- *Contingent Reserves or Contingent Sovereign Capital*.
- *Sovereign Bonds with Special Features*.
- *Diversification and Hedging Related to Management of Foreign Reserves*.
- *Other types of swaps: assets, equity, or debt.*

The examples of the assets selection in the modern *investment portfolio*, created in the *Financial Toolbox(TM)* in the ***Matlab***, are shown in Figs. 4-11. Specifically, the examples show how to set up and optimize the *mean-variance portfolio* with the particular focus on the ***two-fund theorem***, the impact of ***transaction costs*** and ***turnover constraints***, the maximization of the ***Sharpe ratio***, and the two popular hedge-fund strategies - ***dollar-neutral*** and ***130-30 portfolios*** in *Grinold and Kahn (2000), Markowitz (1952, 1959), Lintner (1965), Sharpe (1966), Tobin (1958), Treynor and Black (1973)*.

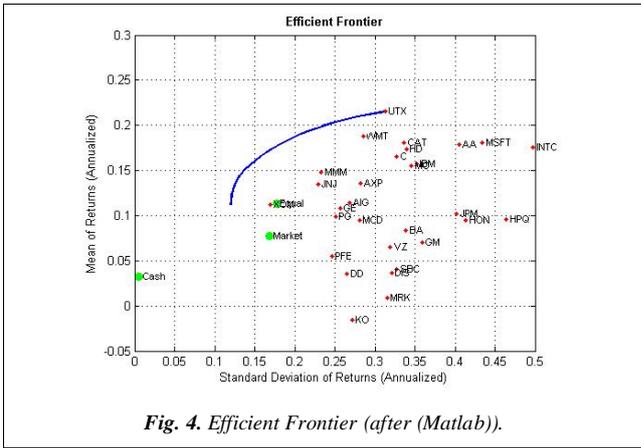

*Fig. 4. Efficient Frontier (after (Matlab)).*

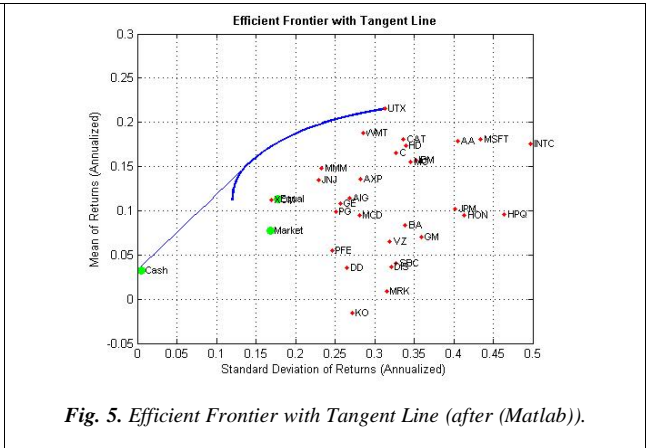

*Fig. 5. Efficient Frontier with Tangent Line (after (Matlab)).*

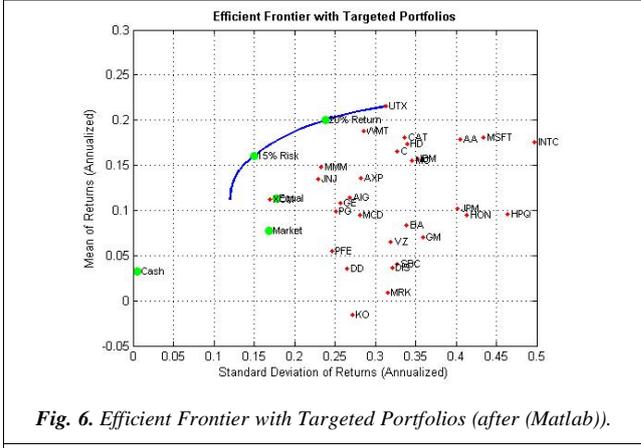

*Fig. 6. Efficient Frontier with Targeted Portfolios (after (Matlab)).*

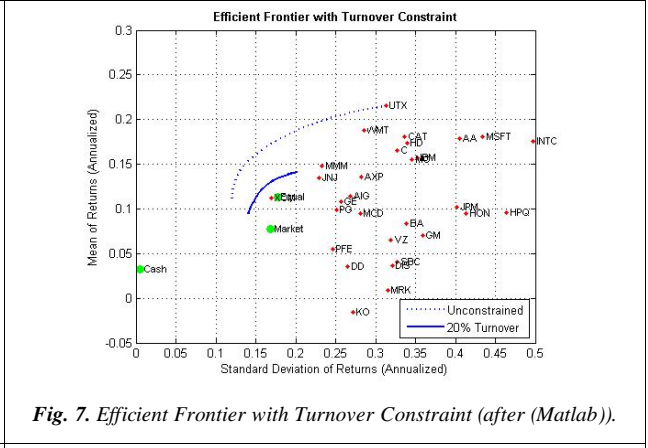

*Fig. 7. Efficient Frontier with Turnover Constraint (after (Matlab)).*

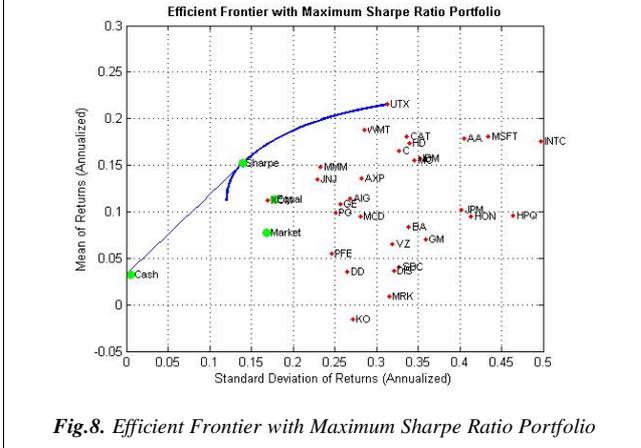

*Fig.8. Efficient Frontier with Maximum Sharpe Ratio Portfolio (after(Matlab)).*

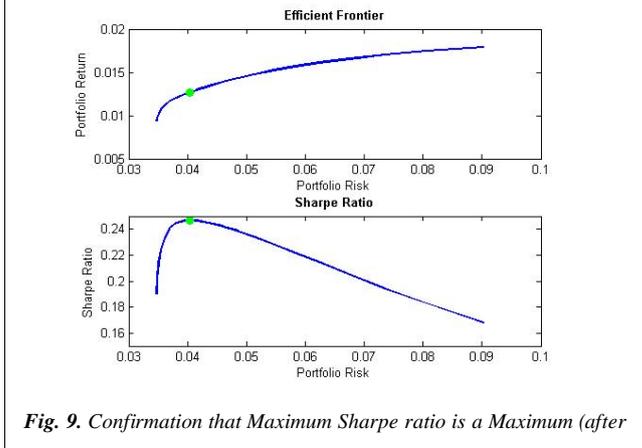

*Fig. 9. Confirmation that Maximum Sharpe ratio is a Maximum (after (Matlab)).*

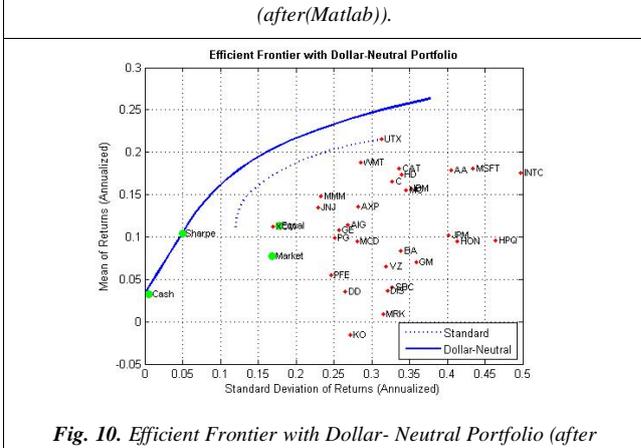

*Fig. 10. Efficient Frontier with Dollar- Neutral Portfolio (after (Matlab)).*

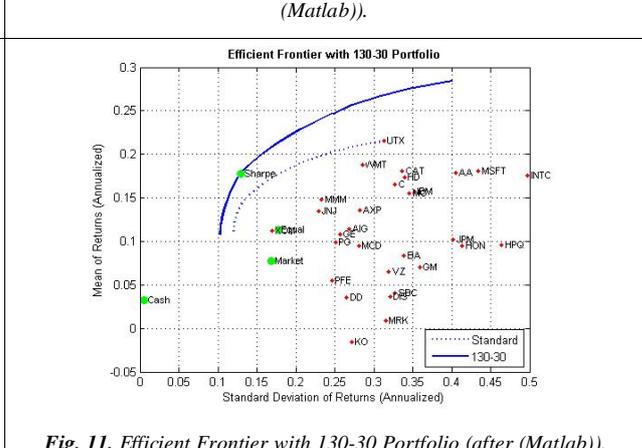

*Fig. 11. Efficient Frontier with 130-30 Portfolio (after (Matlab)).*

***Tab. 1.** Efficient Frontier Modeling in Financial Toolbox(TM) in Matlab (after (Matlab 2012)).*

We intend to optimize the allocation of assets in the *investment portfolio*, modeling the stability of the *investment portfolio*, with the use of the *dynamic chaos theory*. *Alexandrov, Khinchin (1953), Gleick (1988)* explain that the theoretical foundations of the *science of chaos* were created by *Kolmogorov (1931, 1937, 1940, 1941, 1959, 1985, 1986)*. A number of scientists, including *Ulam, von Neumann (1941), Sharkovsky (1964, 1965), Li, Yorke (1975), May (1974, 1976), Feigenbaum (1978, 1979, 1980), Lanford (1982), Feigenbaum, Kadanoff, Shenker (1982), Sharkovsky, Maistrenko, Romanenko (1986), Klimontovich (1989, 1990), Anishenko (1990), Anishenko, Vadivasova, Astakhov (1999),* made significant contributions to the development of the *science of chaos*. Let us review the various dynamic regimes of nonlinear dynamic system, which can be characterized with the application of **logistic function** in the frames of *dynamic chaos theory*. *Kuznetsov (2001)* explains that the first attempt to describe the change of population number was made by *Thomas Maltus (1766-1834):* $x_{n+1} = Rx_n$, where *R* is the parameter, defining the life conditions of population. *Kuznetsov (2001)* clarifies that the population growth can be modeled by the introduction of the nonlinear quadratic term in the equation:
$$x_{n+1} = r(x_n - x_n^2).$$

The graphics of the logistic function *f(x)* is shown in Fig. 12.

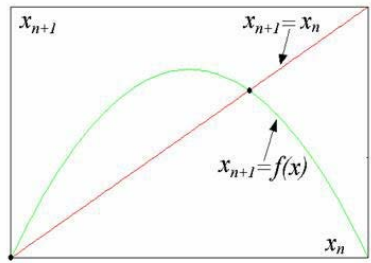

*Fig. 12. The graphics of logistic function (after Saratov group of theoretical nonlinear dynamics (2012)).*

The two fixed stable points are shown on the graphic. The first point is $x_0=0$. The stability of point can be researched by calculating the derivative: $f'(x) = (r - 2rx)|_{x=0} = r.$

The point *r* is non-stable at *r>1*, and its movement dynamics is shown in Fig 13.

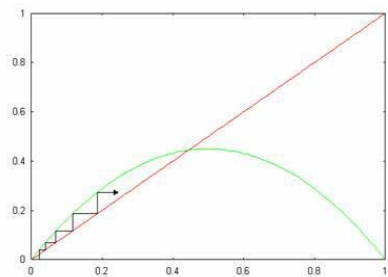

*Fig. 13. The graphics of movement dynamics of non-stable point r (after Saratov group of theoretical nonlinear dynamics (2012)).*

All the iterations of non-stable point *r* at *r>1* converge to the second fixed stable point $x_0$, which is shown in Fig. 14

$$x_0 = rx_0(1 - x_0), \quad x_0^* = 1 - \frac{1}{r}.$$

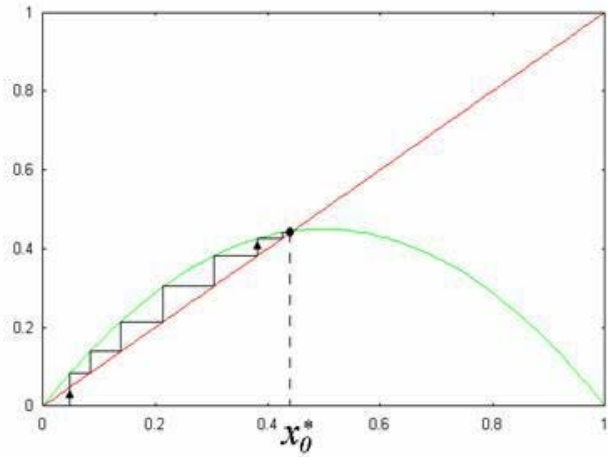

***Fig. 14.*** *The graphics of movement dynamics of non-stable point r with multiple iterations, which converge to point $x_0$ (after Saratov group of theoretical nonlinear dynamics (2012)).*

The point $x_0^*$ is non-stable at *r>3*, and it has the movement dynamics, which is shown on the iteration diagram in Fig 15

$$\mu = f'(x_0^*) = r - 2rx_0^* = r - 2r\left(1 - \frac{1}{r}\right) = 2 - r,$$

$$\mu = |f'(x_0^*)| > 1.$$

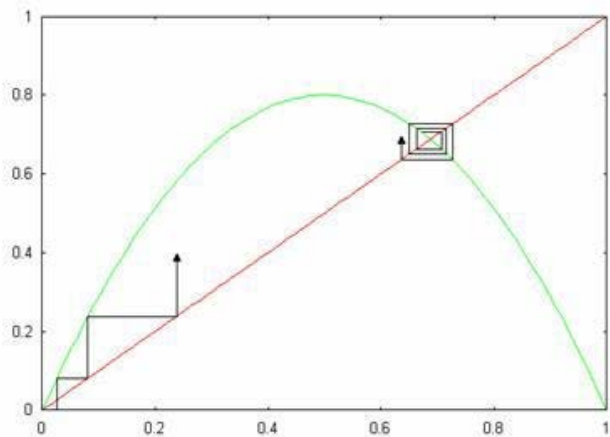

***Fig. 15.*** *The graphics of movement dynamics of non-stable point $x_0^*$ with multiple iterations, which converge to point $x_0$ (after Saratov group of theoretical nonlinear dynamics (2012)).*

At $r=3$, the point $x_0^*$ has the movement dynamics, which is shown on the iteration diagram in Fig 16.

$$\begin{cases} f(x_2) = x_1 \\ f(x_1) = x_2 \end{cases} \Rightarrow \begin{cases} x_1 = rx_2(1-x_2) \\ x_2 = rx_1(1-x_1) \end{cases}$$

$$\begin{cases} x_1 - x_2 = (x_1 - x_2)[r(x_1 - x_2) - r] \\ (x_1 + x_2)^2 - 2x_1 x_2 = (x_1 + x_2)\left(\dfrac{r-1}{r}\right) \end{cases} \Rightarrow \begin{cases} x_1 + x_2 = \dfrac{1+r}{r} \\ x_1 x_2 = \dfrac{1+r}{r^2} \end{cases}.$$

$$x_{1,2} = \frac{1+r \pm \sqrt{(r-3)(r+1)}}{2r}$$

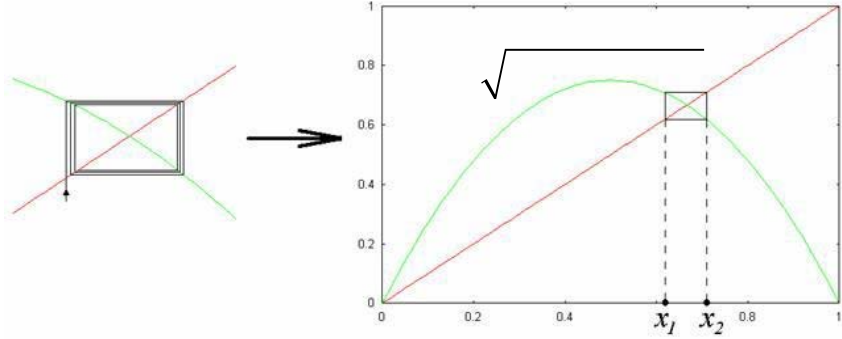

***Fig. 16.*** *The graphics of movement dynamics of non-stable point $x_0^*$ with multiple iterations, which converge to point $x_0$ (after Saratov group of theoretical nonlinear dynamics (2012)).*

In the case of the period two cycle at $3<r<3,499$, the following expression is true

$$x_1 = f(x_2) = f(f(x_1)).$$

Let us find the criteria of cycle stability, showing that the period four cycle can be realized at $r=3,499$

$$\mu = [f(f(x_1))]' = f'(f(x_1))f'(x_1) = f'(x_2)f'(x_1).$$
$$f'(x) = r - 2rx$$
$$\mu = r^2(1-2x_1)(1-2x_2) = r^2[1 - 2(x_1 + x_2) + 4x_1 x_2] =$$
$$= r^2\left[1 - 2\frac{1+r}{r} + 4\frac{1+r}{r^2}\right] = -r^2 + 2r + 4.$$
$$r = 1 + \sqrt{6} = 3,499.$$

Presenting the map of dynamic regimes, *Kuznetsov (2001)* considers the following *logistic equation*

$$x_{n+1} = 1 - \lambda x_n^2$$

where $\lambda$ is some parameter, presenting the solutions of *logistic equation* on the map with the different dynamic regimes in Fig 17.

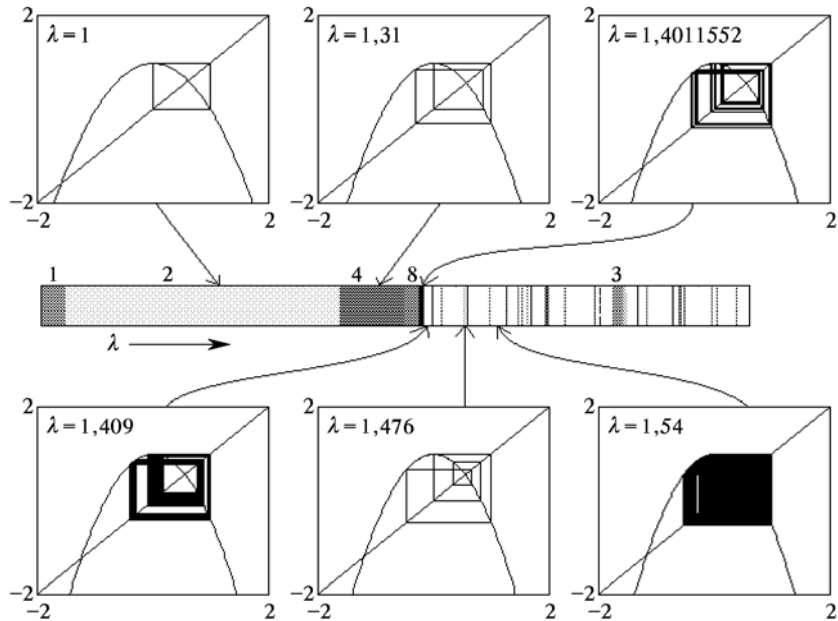

***Fig. 17.** Map of dynamic regimes. Regions with different regimes on axis of parameter $\lambda$ in logistic equation are shown (after Kuznetsov (2001)).*

It can be seen that the period 2, 4, 8, 16 cycles appear in the simple nonlinear dynamic system at the increase of parameter *r*, resulting in the origination of dynamic chaos.

In Fig. 18, the corresponding ***bifurcation diagram*** is also shown in *Kuznetsov (2001)*.

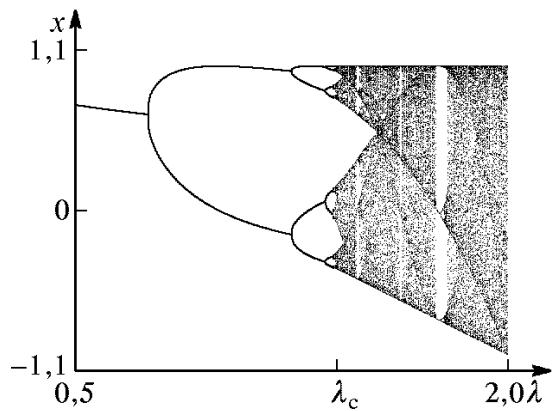

***Fig. 18.** Bifurcation diagram (after Kuznetsov (2001)).*

In Figs. 19-27, *Medvedeva (2000)* modeled the *dynamics of logistic function*, considering the behaviour of a simple nonlinear dynamic system, which can be described by the logistic function $f_\lambda = \lambda x(1-x)$.

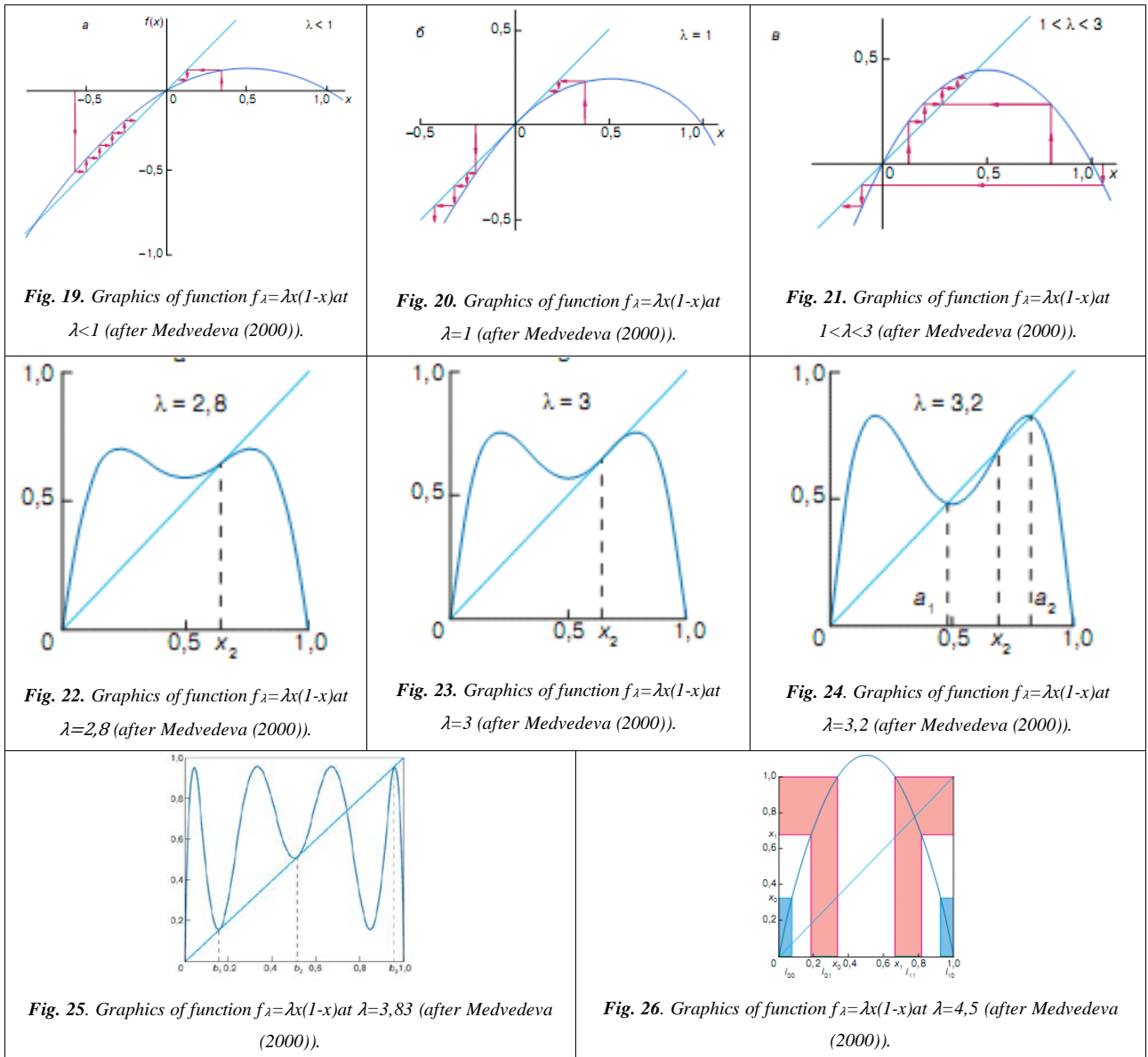

**Tab. 2.** *Graphics of function $f_\lambda = \lambda x(1-x)$ at various parameters $\lambda$ (after Medvedeva (2000)).*

At $\lambda=3,569$, the transition to the chaos is observed as shown in Fig. 27 in *Medvedeva (2000)*.

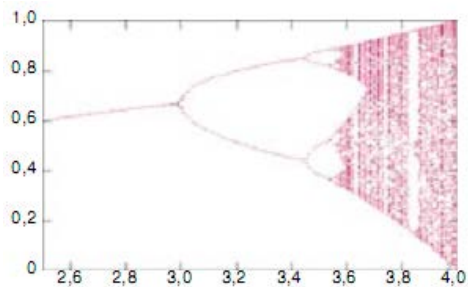

**Fig. 27.** *Bifurcation diagram of function $f_\lambda = \lambda x(1-x)$, created by its iterations from $100^{th}$ to $400^{th}$ (after Medvedeva (2000)).*

The *dynamic chaos* was a subject of intensive research in the economics and finances as explained in *Hsieh (1991)*. *Peters (1996)* reviewed the chaos and order in the capital markets, demonstrating that the simple nonlinear dynamic systems can be characterized by the logistic equation

$$P_{t+1} = aP_t(1-P_t)$$

In Fig. 28, the *bifurcation diagram, created with the help of logistic equation is shown in Peters (1996)*.

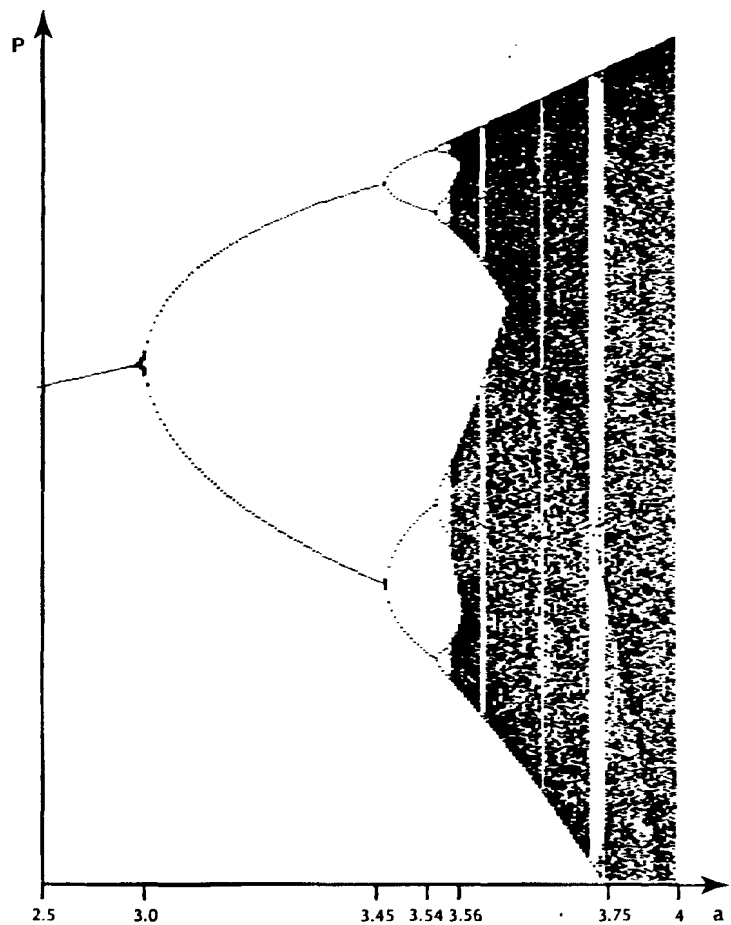

*Fig. 28. Bifurcation diagram: logistic equation (after Peters (1996)).*

*Peters (1996)* writes that the *Feigenbaum* found the following dependence in the alternation of bifurcations

$$\frac{(b_n - b_{n-1})}{(b_{n+1} - b_n)} = 4.669201609,$$

where *b* is the value of parameter *a* at the bifurcation number *n*; *F=4.669201609* is the ***Feigenbaum number***.

Most recently, *Shiryaev (1998)* reviewed the **nonlinear chaotic models**, highlighting a well known fact that the economic and financial systems can be characterized as the *chaotic systems* or the *deterministic nonlinear systems*. *Shiryaev (1998)* considers the *nonlinear dynamic system*, described by the following **logistic equation**

$$x_n = \lambda x_{n-1}(1 - x_{n-1}), \quad n \geq 1, \quad 0 < x_0 < 1,$$

demonstrating that the nonlinear dynamic system has a number of the stable and unstable states at the increase of parameter $\lambda$, resulting in the transition to the chaos state at the parameter $\lambda=3,6$. *Shiryaev (1998)* notes that the following expression is true in the case of all the parabolic systems

$$\frac{\lambda_k - \lambda_{k-1}}{\lambda_{k+1} - \lambda_k} \to F, \quad k \to \infty,$$

where *F = 4.669201* is the **Feigenbaum number** in *Lanford (1982)*. In Figs. 29-34, the illustrations of *bifurcation diagram* are shown in *Shiryaev (1998)*.

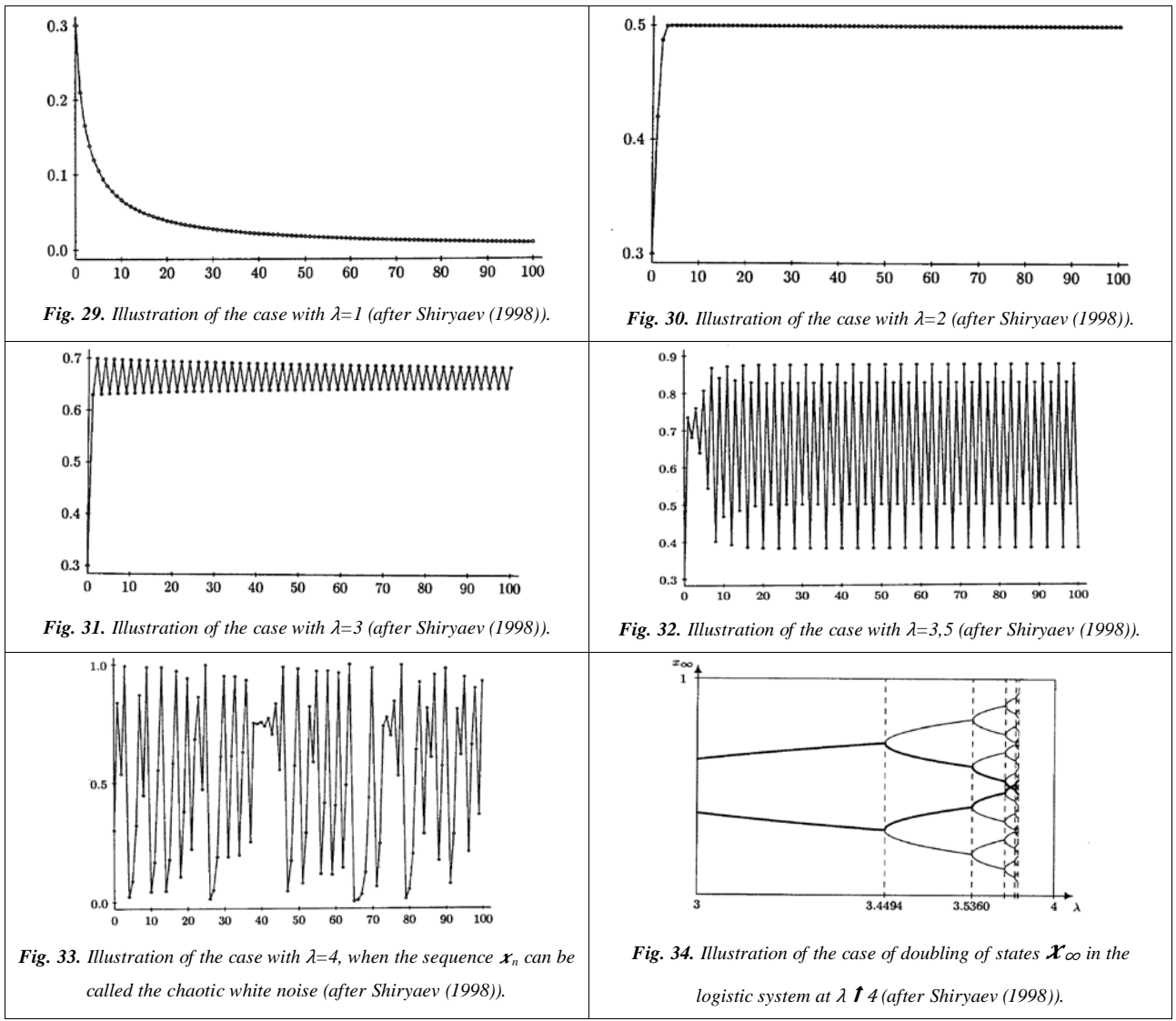

*Fig. 29.* Illustration of the case with $\lambda=1$ (after Shiryaev (1998)).

*Fig. 30.* Illustration of the case with $\lambda=2$ (after Shiryaev (1998)).

*Fig. 31.* Illustration of the case with $\lambda=3$ (after Shiryaev (1998)).

*Fig. 32.* Illustration of the case with $\lambda=3,5$ (after Shiryaev (1998)).

*Fig. 33.* Illustration of the case with $\lambda=4$, when the sequence $x_n$ can be called the chaotic white noise (after Shiryaev (1998)).

*Fig. 34.* Illustration of the case of doubling of states $x_\infty$ in the logistic system at $\lambda \uparrow 4$ (after Shiryaev (1998)).

***Tab. 3.*** *Illustration of bifurcation diagram (after Shiryaev (1998)).*

Going to the detailed research discussion, let us emphasis that the *geometrical approach* in the modeling of the complex physical behavior of the nonlinear dynamic systems is a well known mathematic method, which is usually applied to make the accurate characterization of the nonlinear dynamic properties of the nonlinear dynamic systems in *Klimontovich (1982, 1990)*; *Anishenko (1990, 1999, 2000)*; *Mosekilde (1996, 1996-1997)*; *Kuznetsov (1996-1997, 2001)*; *Ledenyov D O, Ledenyov V O (2012); Nikulchev (2007, 2011)*. The *geometrical approach* is also applied to the modeling of the complex physical behaviors of the nonlinear dynamic systems in the economics and finances in *Peters (1996), Mosekilde (1996)*, *Shiryaev (1998), Ilinski (2001)*, *Smirnov (2010)*, however the application of this advanced econophysical technique in the finances is still relatively limited, because of a number of reasons. In this research paper, we apply the *learning analytics* in *DeGioia (2012)*, *Faust (2012)*, *Rubenstein (2012)* together with the ***integrative creative imperative intelligent conceptual co-lateral adaptive thinking*** in *Martin (1998-1999, 2005-2006, 2007, 2008)* with the purpose to advance our knowledge on the modeling of an *investment portfolio* with the diversified assets in the conditions of the *nonlinear dynamical financial system*. In our representation, we assume that the real financial system is a *nonlinear dynamic financial system*. Therefore, we argue that, in the conditions of the *nonlinear dynamic financial system*, it is not possible to make the accurate characterization of *the combining risky investments* in *the investment portfolio* with the use of the *efficient frontier*, representing the dependence of *the expected return* on *the standard deviation of return* at the given moment of time, in *Markovitz (1952)*. ***We propose to use the dynamic regimes modeling on the bifurcation diagram, based on the dynamic chaos theory, with the purpose to make the accurate characterization of the dynamic properties of the combining risky investments in the investment portfolio, namely to precisely characterize the stability of investment portfolio.*** Thus, in our opinion, the more accurate characterization of *the investment portfolio* can certainly be obtained by taking to the account the existing *nonlinearities* in the *nonlinear dynamic financial system* and by considering the dynamic behavior of the *combining risky investments* in the *investment portfolio*. Therefore, we propose that, at the first approximation, the dynamic behavior of the *combining risky investments* in the *investment portfolio* has to be closely approximated by ***a simple nonlinear system model*** with the physical characteristics, which strongly depend on the initial conditions of *the nonlinear dynamic financial system*. The physical behavior of a *simple nonlinear system* can be modeled, applying the ***logistic function*** in *May (1974, 1976), May, Oster (1976), Gleick (1988), Mosekilde (1996), Shiryaev (1998), Medvedeva (2000), Kuznetsov (2001)*. The computed solutions of *logistic equation* can be presented on the ***bifurcation diagram***, where *the expected return* is plotted at the axis *Y* and the parameter *λ, which has a complex dependence on the forcing amplitude of the standard deviation of return* is plotted at the axis *X*. In our case, the creation of the *bifurcation diagram* to characterize the dynamic behavior of *the combining risky investments* in *the investment portfolio* includes the plotting of the values of *the logistic function $f_\lambda^n(x_0)$* at the vertical axis *Y* and the values of the parameter *λ* at the horizontal axis *X* after the computing of the *1000* iterations. In Fig. 35, the *3D bifurcation diagram* shows the transition of the *combining risky investments* in the *investment portfolio* from ***the stable state*** to ***the chaotic state***, which is realized through ***the period doubling bifurcations*** by which the *1 : 2* mode locking solution is transformed into the *2 : 4*, *4 : 8*, *8 : 16* solutions. It can be seen that, at the *high enough forcing amplitudes* of *the standard deviation of return*, the nonlinearities start to appear and the model begins to bifurcate, exhibiting the alterations between the high maximum and the low maximum. ***The period two bifurcation*** occurs at the parameter *λ≈3*. ***The period four bifurcation*** appears at the parameter *λ≈3,45*. At the further increase of the forcing amplitudes of *the standard deviation of return*, the cascade of bifurcations originates, and the model of simple nonlinear system transits to the state of ***chaos*** with no regular periods at the parameter *$λ_∞=3,5699$*. We researched the complex dynamics of *the logistic function* up to the parameter *λ≈4*, obtaining the research results, which are in a good agreement with the ***Sharkovsky-Yorke theorem*** in *Sharkovsky (1964, 1965, 1986)*; *Li, Yorke (1975)*. We would like to note that the sharp expansion of the ***chaotic attractor*** is registered at the big enough forcing amplitudes of *the standard deviation of return*, resulting in a *crisis*, this research result complies with the research findings in *Grebogi, Ott, Yorke (1982)*. We would like to comment that, in our consideration, the choice of *the logistic equation* to describe a model of the *simple nonlinear system* was determined by the fact that this is a relatively well known mathematical approach to represent the complex behavior of a simple nonlinear system. There are many other nonlinear differential equations, which can also be researched in *Strogatz (1994)*, *Sauer et al (1996)*, *Bunde, Havlin (2009)* in the quantum economics in *Maslov (2006)*. In addition, we would like to add that there are various scenarios of stability disappearance in *the simple nonlinear system*, resulting in its transition from the *stable state* to the *chaotic state* as explained in *Kuznetsov (2002)*. In our case, we consider the chaos transition scenario by the means of the *period doubling bifurcations* only. In addition, it may be interesting to note that we also analyze the existing possibilities and limitations in the *Lyapunov analysis*, which can be applied with the purpose to precisely characterize the chaos with the *Lyapunov exponents*. For example, we propose the algorithm to compute the stability of the combining risky investments in the *investment portfolio*, which is based on *the Ledenyov investment portfolio theorem.*

***Ledenyov investment portfolio theorem:*** *The investment portfolio is stable in the case, when any pair of randomly selected assets from the investment portfolio is stable, satisfying the Lyapunov stability criteria; namely the two randomly selected assets must have the two close trajectories at the start and continue to have the two close trajectories always.*

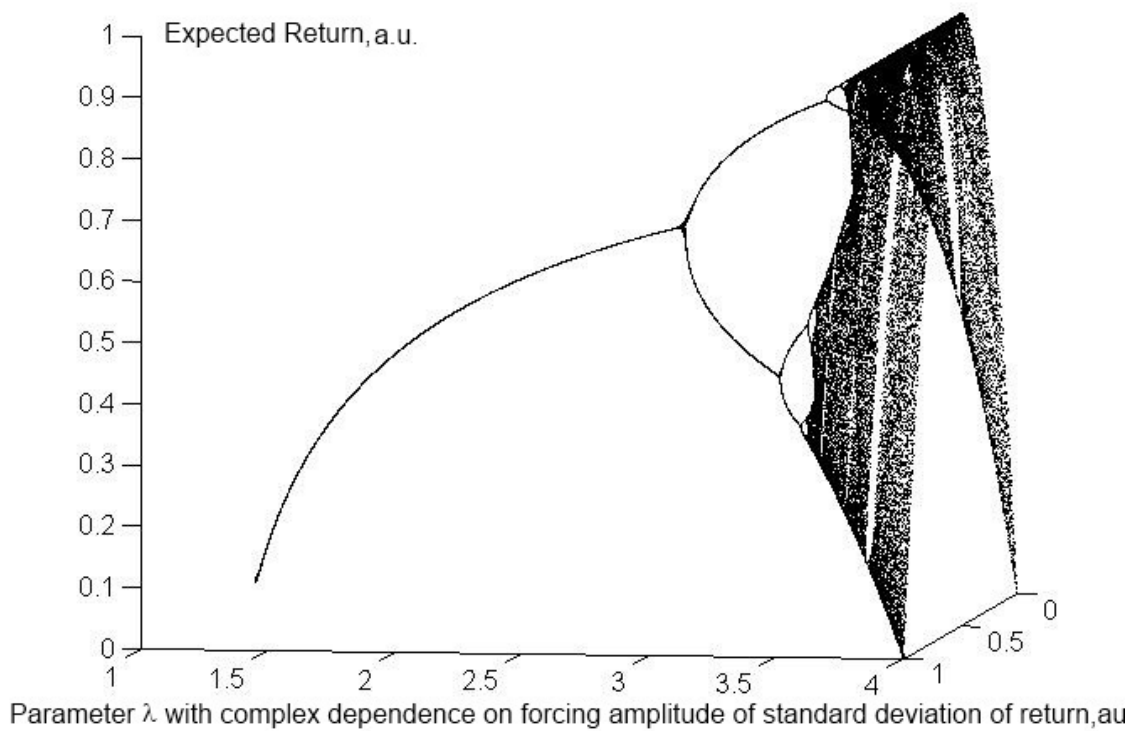

*Fig. 35. 3D Bifurcation diagram for accurate characterization of dynamic properties of combining risky investments in investment portfolio in nonlinear dynamic financial system (Ledenyov D O, Ledenyov V O (2012) Software in Matlab R2012).*

In this research article, we highlight the fact that the investment economy is a main characteristic of prosperous society. We emphasis the important observed feature that a biggest part of investment capital is mostly invested in the research and innovation in the investment economy in *Sperling (2012)*. We also note that the management of the *investment portfolio* is a difficult practical task, which is usually solved by the investment, commercial and central banks with the application of *modern portfolio theory* in the conditions of *investment economy*. We used the *learning analytics* together with the *integrative creative imperative intelligent conceptual co-lateral adaptive thinking* with the purpose to advance our scientific knowledge on the diversified investment portfolio building in the *nonlinear dynamic financial system*. We applied the *econophysics* principles and the *econometrics* methods with the aim to find the solution to the problem of the optimal allocation of assets in the *investment portfolio*, using the advanced risk management techniques with the efficient frontier modeling in agreement with the modern portfolio theory and using the stability management techniques with the dynamic regimes modeling on the bifurcation diagram in agreement with the dynamic chaos theory. We showed that the bifurcation diagram, created with the use of the logistic function in *Matlab*, can provide some valuable information on the *stability* of *combining risky investments* in the *investment portfolio*, helping to optimize the assets allocation in the *investment portfolio*. We propose the *Ledenyov investment portfolio theorem*, based on the *Lyapunov stability criteria*, with the aim to create the optimized investment portfolio with the uncorrelated diversified assets, which can deliver the increased expected returns to the institutional and private investors at the acceptable levels of risk in the nonlinear dynamic financial system in the frames of investment economy in *Justin Lin (2012)*.

Authors are very grateful to the *Graduate School of Economics and Business Administration at Hokkaido University*, Sapporo, Hokkaido, Japan for giving us a wonderful opportunity to conduct the research on the highly innovative research papers, written by the *Japanese* scientists. We appreciate Prof. Geoffrey G. Jones from the *Harvard Business School Harvard University* in the *USA* for the thoughtful discussion on the origin of chaos in the finances during our memorable meeting at the *Munk Centre for International Studies, Trinity College, University of Toronto* in Ontario, Canada in 2006. We also thank the *Harvard Business School, Harvard University* in Boston, *USA* for a presented opportunity to make a detailed analytical study on the current state of the investment portfolios research in the finances, using an electronic collection of working papers at the *HBS*. The first author appreciates Prof. Janina E. Mazierska, *Electrical and Computer Engineering Department, School of Engineering and Physical Sciences, James Cook University*, Australia for an opportunity to make the advanced innovative research on the modeling of nonlinear dynamic microwave resonant systems in the field of superconducting electronics during more than *12* years. The second author would like to thank Prof. Erik Mosekilde from the *Center for the Modeling, Nonlinear Dynamics and Irreversible Thermodynamics* at the *Technical University of Denmark* in Lyngby in Denmark for the numerous thoughtful long discussions on the nonlinear dynamics, including the nature of chaos and hyper-chaos in the financial, economic and managerial systems in *1996-1997*. Prof. Harvey R. Campbell from the *Fuqua Business School* is appreciated for an opportunity to discuss the financial terminology in *Morgenson, Campbell (2002)*, *Campbell (2005-2006)*, and our thanks are also given to a group of leading professors from the *Fuqua Business School, Duke University* for the discussions on the application of different mathematical distributions in the finances during our numerous business meetings in Durham, North Carolina in the USA in 2005. The second author is grateful to Profs. Roger L. Martin and John C. Hull from the *Rotman School of Management, University of Toronto* in Toronto, Canada for a wonderful opportunity to learn more about the integrative thinking, derivatives, and risk management in the finances in North America at the *Rotman School of Management, University of Toronto* in Toronto, Canada in *1998-1999* and in *2005-2006*. The second author thanks Prof. Robert F. Engle III, *New York University* for the thoughtful scientific discussion on the modern portfolio, risk management and nonlinear dynamic chaos theories during our memorable meeting at the *Rotman School of Management, University of Toronto* in Toronto, Canada in 2006. The second author is grateful to Martin Wolf, *Chief Economic Commentator*, *Financial Times* for the insightful discussion on the globalization during our business meeting at the *Rotman School of Management, University of Toronto* in Toronto, Canada in 2006. Prof. Joseph Stiglitz*, Columbia University* is greatly acknowledged for the interesting debate on the topics of globalization and its discontents in 2002. Finally, Lionel Barber, *Editor-in-Chief, Financial Times* is appreciated for the discussions on the financial topics with more than one hundred global leaders, economists, financiers in the *Financial Times* in London in the *UK* in recent years.

*E-mail: dimitri.ledenyov@my.jcu.edu.au
References:
**1**. Bachelier L 1900 Theorie de la speculation *Annales de l'Ecole Normale Superieure* vol 17 pp 21-86.
**2.** Ueda 1904 Shogyo Dai Jiten (The Dictionary of Commerce) Japan.
**3**. Ueda 1937 Keieikeizaigaku Saran (The Science of Business Administration, Allgemeine Betriebswirtschaftslehre) Japan.
**4**. Mano O 1968-1969 On the Science of Business Administration in Japan *Hokudai Economic Papers* vol **1** pp 77-93.
**5**. Mano O 1970 The Development of the Science of Business Administration in Japan since 1955 *Hokudai Economic Papers* vol **2** pp 30-42.
**6**. Merton R C 2001 Future Possibilities in Finance Theory and Finance Practice Working Paper 01-030 Harvard Business School Harvard University Boston USA pp 1-43.
**7.** Mantegna R N, Stanley H E 1999 Introduction to Econophysics *Cambridge University Press* Cambridge UK.
25